\documentclass[runningheads,a4paper,oribibl]{llncs}

\usepackage[T1]{fontenc} 
\usepackage{lmodern}
\usepackage{newtxmath}
\usepackage{bm} 
\usepackage[table]{xcolor}
\usepackage[pdftex]{graphicx}
\usepackage{amsmath}
\usepackage{amssymb}
\usepackage{amsfonts}
\usepackage{url}
\usepackage[caption=false,font=footnotesize]{subfig}
\usepackage{cspsymb}
\usepackage{fancyvrb}
\usepackage{multirow}
\usepackage{color}
\usepackage{colortbl}
\usepackage[utf8]{inputenc}
\usepackage{tabularx}
\usepackage{array}
\usepackage{eucal}
\usepackage{stfloats}
\usepackage[noadjust]{cite}
\usepackage{listings}

\definecolor{keycolor}{rgb}{0.5,0,0.33}
\definecolor{staticcolor}{rgb}{0.167,0,1} 
\definecolor{strcolor}{rgb}{0.167,0,1}
\definecolor{commentcolor}{rgb}{0.25,0.5,0.375}

\renewcommand{\bf}{\textbf}

\newcommand\encrypt[2]{\mathsf{E}_{pk_{#1}}(#2)} 

\newcommand\sign[2]{\mathsf{S}_{sk_{#1}}(#2)}

\newcommand\ie{i.e.\ }

\newcommand\etal{\emph{et al.}}

\newcommand{\Star}{STAR-Vote\ }

\newcommand{\rom}[1]{\uppercase\expandafter{\romannumeral #1\relax}}

\urldef{\mailsa}\path|{muratmoran, dwallach}@gmail.com|
\newcommand{\keywords}[1]{\par\addvspace\baselineskip
\noindent\keywordname\enspace\ignorespaces#1}

\begin{document}

\mainmatter
%
\title{Verification of STAR-Vote and Evaluation of FDR and ProVerif}
\titlerunning{Verification of STAR-Vote}

\author{Murat Moran \and Dan S. Wallach}

\institute{William Marsh Rice University, \\
Computer Science Department\\
6100 Main St, Houston, TX 77005, USA\\
\mailsa\\
\url{www.cs.rice.edu}
}

\maketitle

\begin{abstract}
  We present the first automated privacy analysis of STAR-Vote, a real
  world voting system design with sophisticated ``end-to-end''
  cryptography, using FDR and ProVerif. We also evaluate the
  effectiveness of these tools.  Despite the complexity of the voting
  system, we were able to verify that our abstracted formal model of
  STAR-Vote provides ballot-secrecy using both formal
  approaches. Notably, ProVerif is radically faster than FDR, making
  it more suitable for rapid iteration and refinement of the formal
  model.
\end{abstract}
 
\section{Introduction}
\label{intro}

Security systems employ protocols to ensure their desired goals over a hostile network such as that the communication between agents is authenticated and/or the information that needs to be confidential is indeed confidential. They also aim to provide integrity, key distribution, non-repudiation, and other such properties. However, they are always a target for some malicious activity. Moreover, as the complexity of security-critical systems has grown, rigorous verification and secure implementation gains importance. In our case, cryptographic voting systems have multiple actors exchanging messages, to achieve a variety of important goals, requiring a careful system analysis to ensure there isn't a subtle problem. Formal methods has been shown to be a well suited methodology for analysis of cryptographic protocols, including famous results such as Lowe's attack~\cite{Low95} on the Needham-Schroeder public-key protocol (NSPK)~\cite{NS78}. Since then, formal methodologies have been applied in the analysis of a variety of cryptographic protocols, and also for electronic voting systems, using automated tools including FDR~\cite{MHS12, MH13}, ProVerif~\cite{BHM08, DKR09, Smyth2011}, Active Knowledge in Security Protocols (AKISS)~\cite{CCK12}, AVISPA~\cite{ABB+05}, TA4SL~\cite{BHKO04} and Scyther~\cite{Cre08}.

Formal methods and their tools differ in their approaches to reasoning (e.g., BAN logic, theorem proving, or attack construction). However, all require the user to hold a deep understanding of how these tools work in order to reason about a system and its specification. Even to an experienced user, these tools raise a variety of challenges. Every tool differs in its expressiveness: the capability of a formal language while modeling protocols to capture their specifications. For example, some tools may not support automation of complex cryptographic primitives such as homomorphic encryption, as used by many voting schemes. 
Secondly, all model checking tools suffer from the general problem of state space explosion. Unlike toy security protocols, with only a few messages exchanged, analyzing complex security protocols can require computation exponential in the size of the protocol, exhausting finite computational resources, much less the patience of the user. Furthermore, some verifiers will cap the number of simultaneous adversaries or concurrent runs of the protocol, reducing the computational complexity but also possibly missing real-world vulnerabilities. Lastly, usability of the tools is very crucial. Some tools might merely say that a protocol is ``correct'' without offering a proof. Others might offer a counter-example to demonstrate a vulnerability, but that counter-example might require significant human effort to consider whether it applies or not to the ``real'' protocol.

In this article, we investigate the challenges in engineering automated analysis of complex security protocols, and evaluate FDR and ProVerif protocol verifiers through modeling and analysis of the STAR-Vote~\cite{BBK+12} voting system with respect to ballot-secrecy requirement. We have chosen these two verifiers as these tools are mature, widely accepted, and have been previously used for analysis of comparable systems.

\paragraph{STAR-Vote Overview}
\Star~\cite{BBK+12} is a DRE-style electronic voting system, using human-readable paper and encrypted electronic records. \Star supports homomorphic tallying of votes and non-interactive zero knowledge (NIZK) proofs that ballots are well-formed. Voters have a receipt to verify their votes are counted-as-cast by visiting a block-chain ``public bulletin board'' structure. Similarly, voters can challenge machines to prove that any given encrypted vote is an accurate record of their intent~\cite{benaloh06simple}, but challenged votes are not counted in the tally.  \Star was designed around the requirements of Travis County (Austin), Texas by a collaboration between academics and the county elections staff.  A variety of different cryptographic primitives are specified, including homomorphic encryption, NIZK proofs, and hash chains. Some important messages are passed on paper while others are passed electronically. \Star includes a controller operated by poll workers, multiple voting terminals operated by voters, and a ballot box which queries these machines before it will accept any given printed ballot. Needless to say, \Star is a perfect example of a complex security protocol, and it's valuable to apply formal modeling tools to understand its correctness.

\begin{figure}
\centering
\includegraphics[width=01.00\textwidth]{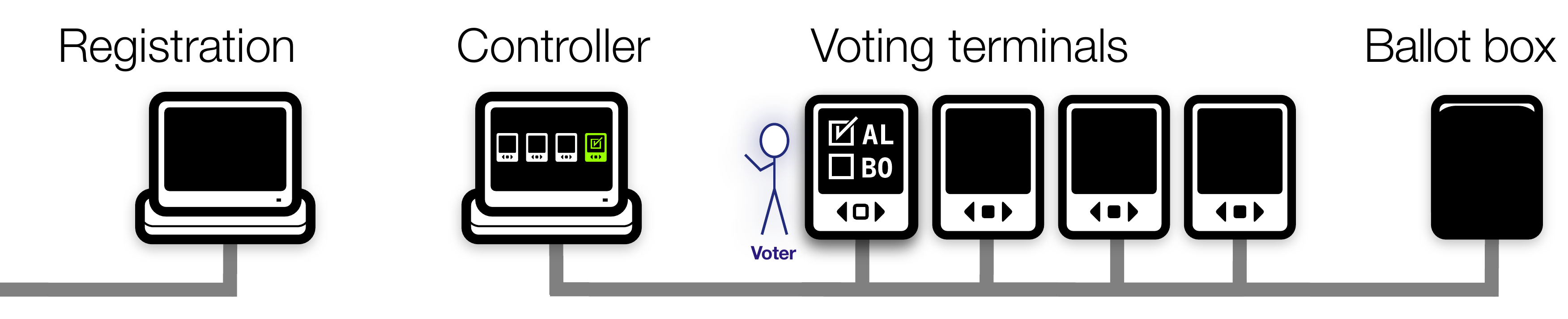}
\caption{A schematic diagram of \Star \label{fig_schematic}}
\end{figure}

From the voter's perspective, all of this complexity is hidden. Figure~\ref{fig_schematic} shows a schematic diagram of the \Star system. An eligible voter goes to a polling station, authenticates to the election official and gets a 1D barcode encoding the voter's precinct and ballot style. This might involve an online database in cases where voters can go to multiple voting centers. \Star maintains an airgap between the voter registration database and the voting system, with the only data that crosses the boundary being the short barcode. The voter then presents the barcode to a poll worker at the controller machine, who scans it to learn the voter's correct precinct and ballot style, and then prints a 5-digit unique code (also called a pin or token). The voter then carries this code to any open voting terminal and is presented with their proper ballot. When complete, the terminal prints a human-readable summary of the voter's choices, which the voter is to deposit in a ballot box, along with a receipt, which the voter can take home. This receipt corresponds to a hash of the ciphertext of the voter's selections which should also later appear on the web bulletin board (hereafter, ``wbb''). The ballot box will refuse to accept anything other than a valid ballot, based on random ballot IDs printed on the ballot and verified with the networked voting machines, thus preventing some ballot stuffing attacks. A voter who recognizes a mistake on the printed ballot can also choose to ``spoil'' that ballot by taking it to the controller rather than depositing it. 

\begin{figure}[!h] 
\centering
\includegraphics[clip, trim=4.5cm 9cm 2cm 1cm, width=0.80\textwidth]{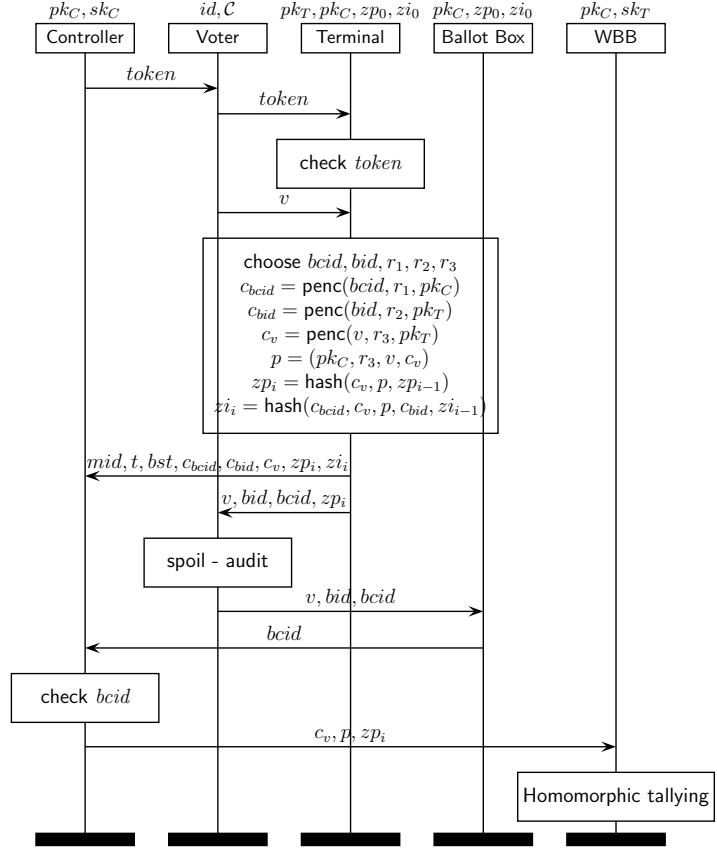} 
\caption{A Simplified \Star voting procedure, where $pk_C$ is controller's and $pk_T$ is trustee's public key, {\it bid} and {\it bcid} are the ballot and ballot cast identifiers respectively, $r$, $zp_0$, and $zi_0$ are random seeds, and {\it mid} is voting terminal id.\label{fig_star}}
\end{figure}

Needless to say, modeling the paper and electronic flow of messages in \Star is a complex task. Figure~\ref{fig_star} illustrates a simplified \Star voting procedure capturing cryptographic message flow on the network. We will discuss the meanings of the relevant messages in the protocol as they arise in our analysis, but even to a quick glance, \Star is sufficiently complicated that we would expect to find challenges in its verification.


\section{STAR-Vote in Communicating Sequential Processes and FDR}
\label{star_fdr}

The STAR-Vote Communicating Sequential Processes (CSP) model is explained in the followings~\footnote{the complete machine readable CSP model is provided at \url{https://muratmoran.com/publications/}}. 

\subsection{STAR-Vote CSP Model}
Our STAR-Vote CSP model consists of five processes running in parallel: a controller process that controls the internal network, a voter process, a ballot marking device (voting terminal) process, an electronic ballot box process that can scan barcodes and sends scanned ballot cast ids to the controller, and a wbb process. Each CSP process has been modeled in a way that they all behave honestly and follow the process flow. As an example, the honest voter process behavior is modeled as the following:

\begin{tabbing}			
$\it{Vot}$\=$\it{er}(id)\defs$\\
\>\;$	\underset{ \begin{subarray}{l}
                \\ c \in \mathcal{C}
								 \end{subarray}}{\Intchoice}
             \begin{pmatrix}\begin{array}{l} 
								choose.id.c \then \\
								scomm.id.Term.c \then \\
							\underset{ \begin{subarray}{l}
								 \\ bid := Bids
									\\ bcid := Bcids
									\\ h:= Hashchains
             \end{subarray}}{\Extchoice} 
             \begin{pmatrix}\begin{array}{l}
						scomm.Term.id?(c, bid, bcid) \then \\
						nbcomm.Term.id?h \then \\
    				scomm.id.Box.Term.(c, bid, bcid) \then \\
						closeElection \then STOP
              \end{array}
       \end{pmatrix}
				\end{array}
       \end{pmatrix}
 $
\end{tabbing}

\noindent where $id$ is the voter identity; $c$ is the candidate; $Term$ is the voting terminal; \emph{Box} is the ballot box process; \emph{bid} is the ballot identifier; and \emph{bcid} is the ballot cast identifier.

When defining processes and messages as in the voter process above, we need two different kinds of channel types; secret $scomm$ and public $comm$ channels in order to maintain the secrecy of security-sensitive information. The STAR-Vote voting system model is then described by the parallel composition of the processes that synchronize on common events as the following:

\begin{tabbing}
$System \defs $\=$\, Voters \parallel Box \parallel Controller \parallel Terminal \parallel Wbb$
\end{tabbing}

\subsubsection{Modeling Assumptions}
In symbolic model checking, it is assumed that cryptographic primitives work perfectly. Hence, the system attacks that may be caused by cryptographic algorithms are not covered in our modeling. We treat cryptographic primitives as symbolic operations with the appropriate algebraic properties, such as; public key encryption: $\encrypt{}{m}$ and digital signatures: $\sign{}{m}$. Hence, an asymmetrically encrypted message can only be retrieved with the corresponding secret key.


Specifically for our CSP STAR-Vote model, the model consists of a limited number of honest agents (2 or 3 simultaneous voters, 2 candidates on the ballot). 
We could modify the model with an increased number of voters, candidates, voting terminals and precincts, but each would require a lot more space in the state base of FDR. Moreover, we assume that the voter chooses the candidate that she would like to vote for before the election begins. This allows us to eliminate false positive attacks. Additionally, the voter verification part of the system is omitted, where the voter can spoil and audit encrypted votes, and can check if her receipt appears on the public bulletin board as we focus on solely voter privacy. Furthermore, the real voting system relies on the hash chains of the encrypted votes, which is simplified and modeled as the hash of an encrypted vote and a nonce provided by the authorities. In terms of randomness in choosing nonce-like terms, such as ballot ids and ballot cast ids, randomness is modeled as the non-deterministic choice of terms over a pre-determined set of terms. STAR-Vote employs a threshold mechanism for creating trustee's public key pairs and decryption of homomorphically multiplied ciphertexts. We model this simply as the public key pairs and asymmetric decryption of a ciphertext respectively. Lastly, encrypted votes are submitted and publicized on the wbb along with non-interactive zero-knowledge proofs, which ensures that ballots are well-formed. However, we further assume that all ciphertexts and ballot forms are well-formed.

\subsubsection{Intruder Model}
In our analysis, we employed an active Dolev-Yao intruder~\cite{DY83} model in CSP, as adapted to voting systems in Moran and Heather~\cite{MH13}. The intruder model supports active intruder behavior: interacting with the protocol participants, overhearing communication channels, intercepting and spoofing any messages that the intruder has learned or generated from its prior knowledge. In this model, intruder processes have a set of deductions rules in order to compose and decompose messages, and a set \emph{Ucomms}, as below, defining the unreliable channels on which the intruder can act. For instance, the following rules enable intruders to have access to any channels from and to the voter $v_1$, where the set \emph{comms} is the union of all communication channels.

\begin{tabbing}
$\it{Ucomms} = \Union($\=$\{q.q'.f \mid q.q'.f \leftarrow comms, q \leftarrow  \set{\it{v_1}}, q'\leftarrow  agents \},$\\ 
\>$ \{q.q'.f \mid q.q'.f \leftarrow  comms, q \leftarrow  agents, q'\leftarrow  \set{\it{v_1}} \})$
\end{tabbing}

The system that will be analyzed is the parallel composition of the renamed voting system model \emph{System}, which enables the messages flowing on unreliable channel to be taken or eavesdropped, and the intruder model \emph{Intruder}. The resulting composition \emph{System} $\parallel$ \emph{Intruder} synchronizes on their common events. In the following section, we will define several sets like \emph{Ucomms} for different intruder capabilities.

\subsection{Analysis with FDR}
\label{sec:csp_analysis}

In this section, we present the first automated analysis of STAR-Vote under an active Dolev-Yao intruder model and the anonymity specification given in~\cite{MHS12}. Accordingly, the following trace equivalence should hold in order to verify that the voting system model \emph{System} provides voter anonymity.

\begin{tabbing} 
$\it{System_1} \hide \eset{scomm} \tequiv \it{System_2} \hide \eset{scomm}$
\end{tabbing}

\noindent, where the processes $\it{System_1}$ and $\it{System_2}$ model two different system behavior: in the first $v_1$ votes for $c_1$ and $v_2$ for $c_2$, and in the second the voters vote other way around. Hence, the intruder with the available public information along with its prior knowledge cannot distinguish these two cases, then we say that the voting system does not leak any information that may link a voter to her cast vote. Here, what information is available to the intruder is defined by a set \emph{Ucomms}, and also by masking or hiding the information that the intruder is not supposed to access.

There are numerous kinds of threat scenarios that we can model in this CSP framework by modifying \emph{Ucomms}. Here, we present three of those cases, denoted as DY1, DY2 and {DY3.

\paragraph{DY1}: the intruder can observe only the public channels ($comm$), and not the channels that should be kept confidential, such as; the channels on which crucial ballot information and voters' choice of candidate are transmitted. Hence, we exclude such information from {\it Ucomms} set.

\begin{tabbing}
${\it Ucomms} = comms \hide (comBallots \union commCandidates)$
\end{tabbing}

\paragraph{DY2 (only 1 honest voter)}: the intruder can act as the Dolev-Yao intruder on all the channels except the voter $v_1$'s communication channels---there exists only one honest voter and the rest is dishonest.

\begin{tabbing}
${\it Ucomms} = comms \hide \Union($\=$\{q.q'.f \mid q.q'.f \leftarrow comms, q \leftarrow  \set{\it{v_1}}, q'\leftarrow  \mathcal{A} \},$\\ 
\>$ \{q.q'.f \mid q.q'.f \leftarrow  comms, q \leftarrow  \mathcal{A}, q'\leftarrow  \set{{\it v_1}} \})$
\end{tabbing}

\paragraph{DY3 (only 1 dishonest voter)}: the intruder can act maliciously only on the channels of the voter $v_3$, who is collaborating with the intruder, and observe other public channels---there exist only 1 dishonest and at least 2 honest voters.

\begin{tabbing}
$\it{Ucomms} = \Union($\=$\{q.q'.f \mid q.q'.f \leftarrow comms, q \leftarrow  \set{\it{v_3}}, q'\leftarrow  \mathcal{A} \},$\\ 
\>$ \{q.q'.f \mid q.q'.f \leftarrow  comms, q \leftarrow  \mathcal{A}, q'\leftarrow  \set{\it{v_3}} \})$
\end{tabbing}

Our analysis found that STAR-Vote provides anonymity under active intruder models: DY1 and DY3, and produced privacy attacks against the system under DY2, all of which were as expected. (Why? Because when there's only one honest voter, the other voters can collude to know the subtotals of their own votes, and then infer the votes of the remaining honest voter.) To give an idea about overall verification times for FDR, Table~\ref{tab:vvotevertimes} illustrates the verification times of the automated analysis of STAR-Vote under different intruder capabilities using an average laptop with $\text{Intel}^{\text{\textregistered}}$ $\text{Core}^{\text{TM}}$ i5 CPU 2.40GHz, and 8GB RAM. The longest run took just under three minutes. This is tolerable, but is far from ideal in terms of the engineering cycle time of evolving the model.

\begin{table*} 
\caption{The FDR verification times for STAR-Vote model under different Dolev-Yao capabilities}
\label{tab:vvotevertimes}
\centering
\begin{tabular}{ccc|ccc|ccc}

\multicolumn{3}{c|}{{\textbf{DY1}}} & \multicolumn{3}{c|}{{\textbf{DY2 }}} & \multicolumn{3}{c}{{\textbf{DY3}}}\\\hline

\multicolumn{1}{c}{Refine} & \multicolumn{1}{c}{States} & \multicolumn{1}{c|}{Time} & \multicolumn{1}{c}{Refine} & \multicolumn{1}{c}{States} & \multicolumn{1}{c|}{Time} &
\multicolumn{1}{c}{Refine} & \multicolumn{1}{c}{States} & \multicolumn{1}{c}{Time} \\
$\tick$ & $128,101$ & 21s & X & $26$ & 2m50s & $\tick$ & $95,917$ & 1m39s \\
\end{tabular}
\end{table*} 

We also extended the model with ballot counting mechanism in the wbb process and measured FDR verification times using the same settings as before. As illustrated in Table~\ref{tab:extendedvertimes}, FDR verifies the model for DY1 and DY3 cases, but not DY2 as FDR crashes once 8GB allocated memory runs out in 45 minutes. However, we were able to verify the extended model using a better server with 128GB RAM in 2 hours. Furthermore, when extending STAR-Vote CSP model by including a hash-chain mechanism, which is used in the original system for integrity purposes, even the server with 128GB crashes before producing a result. Hence, automated verification of the STAR-Vote CSP model extended with more components (e.g., pins, hash-chain, and thresholded mechanisms) seems unrealistic.

\begin{table*} 
\caption{The FDR verification times for extended STAR-Vote model with {\em counting} mechanism under different Dolev-Yao capabilities. "`$-$"' means that FDR crashes before producing a result.}
\label{tab:extendedvertimes}
\centering
\begin{tabular}{ccc|ccc|ccc}

\multicolumn{3}{c|}{{\textbf{DY1}}} & \multicolumn{3}{c|}{{\textbf{DY2 }}} & \multicolumn{3}{c}{{\textbf{DY3}}}\\\hline

\multicolumn{1}{c}{Refine} & \multicolumn{1}{c}{States} & \multicolumn{1}{c|}{Time} & \multicolumn{1}{c}{Refine} & \multicolumn{1}{c}{States} & \multicolumn{1}{c|}{Time} &
\multicolumn{1}{c}{Refine} & \multicolumn{1}{c}{States} & \multicolumn{1}{c}{Time} \\
$\tick$ & $1,201,525$ & 23m1s & $-$ & $-$ & $-$ & $\tick$ & $95,917$ & 1m51s \\
\end{tabular}
\end{table*}

\section{STAR-Vote in the Applied Pi Calculus and ProVerif}
\label{star_proverif}

We briefly explain our STAR-Vote applied pi model in the followings~\footnote{The actual applied pi model is provided at \url{https://muratmoran.com/publications/} for brevity}.

\subsection{STAR-Vote Model in Applied Pi}
Similar to the CSP approach, we modeled the STAR-Vote voting system in the applied pi calculus by means of processes that intercommunicate, allowing verification by ProVerif. As with FDR, this process necessarily involves abstracting away some of the details. Initially, we modeled a set of cryptographic primitives $\Sigma$ that are used in STAR-Vote, and it can be defined as the following;

\begin{tabbing}
$\Sigma =$ \{{\sf ok, pk, hash, sign, dec, penc, zkp, checksign, checkzkp}\} 
\end{tabbing}

\noindent
Function {\sf ok} is a constant; {\sf pk}, {\sf hash} and  {\sf checkzkp} are unary functions; {\sf sign} and {\sf dec} are binary functions; {\sf penc}, {\sf zkp} and {\sf checksign} are ternary. Accordingly, we have the following equations: 

\begin{center}
\begin{tabular}{ll}
${\sf dec}({\sf penc}(m, r, {\sf pk}(sk)), sk) = m$ & (E1) \\ 
${\sf checksign}({\sf sign}(sk, m), m, {\sf pk}(sk)) = {\sf ok}$ & (E2)\\ 
${\sf checkzkp}({\sf zkp}({\sf pk(sk)}, r, m, {\sf penc}(m, r, {\sf pk}(sk)))) = {\sf ok}$ \, & (E3) \\ 
\end{tabular}
\end{center}

Equation E1 enables plaintext $m$ to be extracted using the corresponding secret key $sk$. E2 allows digital signatures to be verified with an appropriate public key {\sf pk}(sk). E3 allows non-interactive zero knowledge proof $p$ to be verified.

For the model, we employ two channel types: public and private. The following process $V(c, b, v)$ models a simplified voter process behavior of STAR-Vote in the applied pi calculus, where $v$ is the candidate of choice, $c$ is a private channel between the voter and voting terminal, and $b$ is a private channel between the voter and the ballot box:

\begin{center}
\begin{tabular}{lcll}
$V(c, b, v)$ & ::= & $\bar{c} \langle v \rangle.$ & \,\, voter enters her vote \\ 
& & $c(v', bid, bcid, zp_i).$ & \,\, receives ballot summary and receipt ($zp_i$) \\
& & if $v = v'$ & \,\, checks if chosen candidate on the ballot \\
& & then $\bar{b} \langle (v, bid, bcid) \rangle.$ & \,\, casts her ballot 
\end{tabular}
\end{center}

Likewise, the rest of the processes that comprise the STAR-Vote model (\ie, voting terminal ($T$), ballot box ($B$), controller ($C$) and web bulletin board ($W$) processes) are defined in terms of the grammar and equational theory above. The STAR-Vote pi calculus model $Star(sk_a, sk_c, v)$ is then described as the composition of these processes, and initialized with public-private key pairs $(pk_a, sk_a)$ and $(pk_c, sk_c)$ for election authority and controller respectively. The system then generates fresh seeds $zp_0$ and $zi_0$, and establishes private channels between trusted participants before the election as the following.

\begin{center}
\begin{tabular}{lll}
$Star(sk_a, sk_c, v)$ &::=& \ let $pk_a = pk(sk_a)$ in let $pk_c = pk(sk_c)$ in \\
  & & \ $\nu\, zp_0. \nu\, zi_0.$ \\
	& & \ $\nu\, ch_{VT}. \nu\, ch_{VB}. \nu\, ch_{BC}.$  \\
	& & \ $\nu\, ch_TC. \nu\, ch_{TW}. \nu\, ch_{CW}.$  \\
	& & \ (\\
	& & \ $V(ch_{VT}, ch_{VB}, v) |$  \\
	& & \ $T(ch_{VT}, ch_{TC}, ch_{TW}, pk_a, pk_c, sk_c, zp_0, zi_0) |$ \\	
	& & \ $B(ch_{VB}, ch_{BC}, pk_c, zi_0) |$ \\			 				
	& & \ $C(ch_{BC}, ch_{TC}, ch_{CW}, pk_c, sk_c) |$ \\				
  & & \ $W(ch_{CW}, ch_{TW}, sk_a, pk_c)$\\
	& & \ ) \\
\end{tabular}
\end{center}

\noindent where $\nu$ is the name restriction (i.e., it creates new names).
		
\subsubsection{Modeling Assumptions}
As in many model checking cases, we abstract away some of the properties or components of the voting system that are analyzed due to either state explosion constraints or other limitations of our model checking tools. The most important of them is the homomorphic tallying of encrypted votes in STAR-Vote. ProVerif and FDR are both incapable of verifying homomorphic encryption. Hence, we consider an election scheme where all encrypted votes are published on the bulletin board after the election closed, decrypted individually and counted publicly. The homomorphic tallying ensures that no single vote is decrypted, thus preserving privacy. Using synchronization points in the model, we make sure the intruder does not gain any information that can link encrypted votes with the plaintext equivalence. That is, the wbb waits until all the votes are decrypted, and then publishes all the plaintext votes. Secondly, we abstract away hash chain of encrypted votes used in STAR-Vote as we focus on privacy and not integrity of the election. Lastly, the spoil-audit and risk-limiting audit mechanisms are not reflected in our model, however voters will ensure that each paper ballot reflects their intent.

Are these assumptions and simplifications reasonable? Certainly they leave out security-critical aspects of the STAR-Vote design like the homomorphic tallying. Were an actual election conducted this way, a voter could use their receipt to prove to a third-party how they voted, and thus enable bribery or coercion of their vote. Nonetheless, the use of synchronization points presents a reasonable simulation of the constraints that an adversary might face with regard to attacking voter privacy, at least under the assumption that voters choose not to share their receipts. In a real election, a voter must be unable to compromise their privacy, {\em even if they want to}.

\subsubsection{Intruder Model}
Unlike the CSP approach, ProVerif does not need a separate implementation of an intruder model. ProVerif instead provides a standard Dolev-Yao intruder model, having access to all the public channels, and special functions to perform a number of malicious actions in order to violate voter privacy. That is, he can use anything available in the context. By specifying a private channel as public we can increase intruder's capabilities. Similarly, corrupt system participants such as voters and voting terminals can be modeled easily by either giving away their cryptographic keys or by publishing their private communication channels to the intruder.

\subsection{Analysis with ProVerif}

Vote-privacy (ballot-secrecy) is defined informally as ``no party receives information which would allow them to distinguish one situation from another one in which two voters swap their votes''\cite{DKR09}. Formally, it is defined as:

\begin{definition} 
\label{def:secrecy}
A voting protocol respects vote-privacy if
\begin{center}
$S[V_A\{a/v\} | V_B\{b/v\}] \approx_l S[V_A\{b/v\} | V_B\{a/v\}]$ 
\end{center}
for all possible votes $a$ and $b$.
\end{definition}

Recently, Blanchet and Smyth~\cite{BS16} have proposed an approach, based on barrier synchronization, to fully automate verification of this definition. It is implemented in the latest version of ProVerif, and supports automated verification of observational equivalence. Barrier synchronization ensures that ballot-secrecy holds by swapping outputs of both sides of the observational equivalence. In order to do that, a compiler first annotates barriers with data to be swapped and channels for sending and receiving data; the compiler then translates the biprocesses with annotated barriers into biprocesses without barriers. 

We have used these barriers in our model for instance when describing the wbb process, which receives an individual encrypted vote and decrypts it. Hence, a synchronization point in between these two events is needed so that the order of the communication does not leak any information related to that particular vote. Definition~\ref{def:secrecy} is reflected to ProVerif using a {\sf choice} operator as:

\begin{tabbing}
$Star(sk_a, sk_c, {\sf choice}[a,b])\, |\, Star(sk_a, sk_c, {\sf choice}[b,a])$
\end{tabbing}

\noindent
where $ska$ and $skc$ are authority's and controller's secret credentials, respectively, which are fed into the system, and $a$ and $b$ are candidate names. 

Having described an abstract model of STAR-Vote, we were able to verify that our model satisfies the ballot-secrecy property using the ProVerif protocol verifier. With the same setting as in the CSP approach (the same laptop with 8GB RAM), ProVerif takes around 1.45s in total to verify our pi calculus model of this complex voting system protocol. ProVerif is also able to find possible attacks when, for instance, there exists a corrupt voting terminal and a ballot box by using the compromised information from these entities either by revealing corresponding secret keys or by making private channels public. 

Extending the model in ProVerif is straightforward. We have also managed to verify two extended versions of this model: first one is extended with pins or tokens, which are given to voters by the controller for authorization purposes and then scanned to a voting terminal, and this version is verified in 9.8s; the second one is extended with hash-chain mechanism, which requires two honest voter processes and other system participants processes extended for two voters. ProVerif verifies this extended model in 2.10m using the same laptop.


\section{Evaluation of the Tools: FDR and ProVerif}
\label{sec:evaluation}

In this paper, having analyzed STAR-Vote voting system mechanically with FDR and ProVerif, we now share our experience with these two tools in this section with respect to expressiveness, usability and efficiency. The tools provide different approaches to protocol verification and make verification of complex security protocols easier than hand-proofs, but they may also suffer from the similar problems such as state explosion. We discuss some of the issues we encountered during our analysis in the following categories.

\subsection{Expressiveness}
We came across several inadequacy of the tools in expressing some of the system components, which needed to be abstracted away. For example, neither FDR nor ProVerif can verify homomorphic tallying, or threshold encryption and decryption. FDR furthermore cannot verify non-interactive zero knowledge proofs unlike ProVerif. Similarly, a typical voting system requirement, coercion-resistance, can be defined in CSP~\cite{HS12}, but FDR does not support its mechanical verification. ProVerif can verify this property but we did not make it a focus of our verification efforts.

FDR is very expressive in its support for many different kinds of channels: public or private, blocked or spoofed, all of which can be defined in terms of functions and sets. ProVerif only supports public and private channels. In practice, this expressivity is necessary in FDR, which does not provide an adversary model, while ProVerif provides a Dolev-Yao adversary that does everything we need.


\subsection{Usability}
We found modeling and expressing protocol participants more straightforward with ProVerif than FDR. FDR frequently complains when the network of protocols is too complex to bring together. However they both guide the user well in finding bugs in the specification. FDR offers a sophisticated user interface, the ProBE CSP animator, which enables checking if processes behave as intended.

In terms of producing and interpreting counter-examples during the analysis, ProVerif {\em sometimes} produces a trace that leads to the attack when the verification does not hold; other times, ProVerif only says that a query does not hold, and terminates. Moreover, when ProVerif {\em does} return a counter-example trace, the task for the user to interpret the trace and locate why the attack occurs is often very difficult; we saw some traces that were 3-4 pages long. This also makes it difficult for a ProVerif user to distinguish whether a trace corresponds to a legitimate attack, since ProVerif can sometimes return false-positives; this might seem terrible, but it's essential to how ProVerif gains its runtime performance. On the other hand, FDR always produces a counter-example when there should be one, and tracing back the attack is smooth and straightforward.

In some cases ProVerif verified our model when it should not have, due to some minor, unrecognized bugs in our model, for instance; a type mismatching of functions or creating new names earlier in the model. Hence, it was not simple to find such bugs during modeling and verification, which may deceive an inexperienced user into analyzing incorrect model.

Lastly, consider the case of modeling a new voting system, starting from our existing STAR-Vote models in both FDR and ProVerif. How hard might it be to derive a new voting system model from our existing one? Code reuse would certainly be a valuable feature. We note that the ProVerif pi calculus model for STAR-Vote is around 100 lines of code while the CSP model is around 500 lines of code. This additional complexity in CSP comes largely from having to specify sets that are used to describe system participants and the intruder's behavior. ProVerif wins for having a generic intruder that we don't need to specify.

\subsection{Efficiency}
Verification times vary in FDR depending on the number of participants and whether the verification holds. We saw runtimes as fast 21 second, and we saw crashes which occurred after 17 hours, and to even run that long we had to move to a much larger computer with 128GB RAM. Needless to say, this can make for a frustrating user experience.

Table~\ref{tab:vvotevertimes} displays verification times for FDR for scenarios with 2 or 3 voters, and Table~\ref{tab:extendedvertimes} illustrates verification times for a model extended with pins. When we add an extra tallying mechanism in the model, the DY1 case increases from 21~seconds to 23~minutes with 10 times more states than before, and in the case of DY2 with tallying, FDR crashes after 45~minutes, on a laptop with 8GB RAM, due to lack of memory. We verified this extended version with a bigger server with 128GB. Additionally, when we extended the model further with hash chain mechanism FDR crashes even with the larger server after 17 hours. Generally, the more ability given to intruder, the longer FDR takes to verify. We note that, to make verification more efficient, FDR3 and FDR4 offer multi-core parallelism features. Unfortunately, they don't support testing observational equivalence of complex models where the left hand side of the equivalence requires more than 10 million states.

We found that ProVerif operates very quickly with models of similar complexity to those we used in FDR. Verification took generally less than two seconds to complete, allowing us to rapidly iterate on our models. We verified two extended versions of the model with pins and hash-chain mechanism in 9.8~seconds and 2.10~minutes respectively. Additionally, ProVerif produces generic results {\ie, independent of the number of concurrent participants in the protocols} unlike FDR. This means that our ProVerif proofs give us a stronger assurance of the correctness of our system.

In terms of man-hours, ProVerif can produce false-positive attacks due to its over-approximation~\cite{CSS15}. Hence, dealing with such false positives takes enormous amount of man-hours and effort. However, FDR does not produce such false positives unless an intruder's power is adjusted improperly, but the user-defined intruder model requires careful attention and takes a lot of time to integrate. 

\section{Related Work}
\label{related}

To date, there have been a few attempts to compare automated security protocol verifiers in the literature. C. Meadows~\cite{Mea96} compares the approaches followed in the tools NRL and FDR with the analysis of NSPK, and concludes that two tools are complementary.
Hussain and Seret~\cite{HS06} presents a qualitative comparison between AVISPA and Hermes in terms of their complexity, ease to use and the conceptional differences between approaches (the comparison is not based on experiments). It is stated that Hermes is more suited for simple protocols, on the other hand, AVISPA is better for complex protocols where you would need scalability, flexibility, and precision. Cas J.F. Cremers \etal~\cite{CLN09} first discuss the types of behavior restriction of the models used by the tools; Casper/FDR, ProVerif, Scyther and AVISPA back-end tools. Then, a performance comparison is made considering an analysis of secrecy and authentication properties. This is the only work that compares our chosen tools ProVerif and Casper/FDR~\footnote{Casper is a compiler that translates protocol description into the CSP language, which is then used by FDR.}. However, the properties that we are dealing with in this paper is not considered since ProVerif was not able to check observational equivalence properties then. Dalal \etal~\cite{DSHJ10} compare ProVerif and Scyther tools considering six different security protocols. The definitions of the models presented in the paper are not language specific, but pseudocode. Lafourcade \etal~\cite{LTV09} analyze a number of protocols dealing with algebraic properties like Exclusive-Or and Diffie-Hellman and compare the results from different tools: OFMC, CL-Atse and XOR-ProVerif or DH-ProVerif in terms of efficiency. The properties that were checked are secrecy, authentication and also non-repudiation for one e-auction protocol. Cortier \etal~\cite{CEK+15} proposed a semi-automatic proof of vote privacy using type-based verification and the tool $\text{rF}^\star$, in which security properties and cryptographic functions are modeled in terms of refinement types. More recently, Lafourcade and Puys~\cite{LP15} focus on performance analysis of a number of tools including a ProVerif extension and analysis of 21 cryptographic protocols dealing with Exclusive-Or (xor) and exponentiation properties like Diffie-Hellman (DH). In the analysis, secrecy and authentication properties are considered. The tools have been evaluated in terms of execution time and memory consumption. It is stated that there is not a clear winner, but more recent tools tend to perform better.
\section{Conclusion}
\label{conclusion}
In this paper, we have presented the first automated privacy verification of the STAR-Vote voting system along with an evaluation of two protocol verifiers: FDR and ProVerif. We verified that our STAR-Vote CSP and pi calculus models provide ballot-privacy, validating our previously informal design and providing further trustworthiness.

Throughout our analysis we had a chance to evaluate these two security protocol verifiers with respect to their expressiveness, usability, and effectiveness. In terms of expressiveness, both tools need further research to pursue in automation of cryptographic primitives. 
Regarding usability, FDR offers more with its inbuilt tools to make sure the model behaves as expected, user interface and counter-examples, which are easy to interpret and trace back to what causes the failure. On the other hand, modeling with ProVerif is more straightforward and requires quite less effort than FDR does. About efficiency, ProVerif is very efficient and quite flexible in modeling and analyzing such complex systems despite the false-positives, which require a special attention. FDR with lazy spy intruder model is neither efficient nor scalable when analyzing such complex systems.

Overall, formal verification helps us understand how complex security protocols work and facilitate their analysis. However, it is still expensive in the sense that it requires a deep understanding of verification tools, experience and a huge amount of human effort. 

Our future work will concentrate on improving the protocol verifiers ProVerif and FDR by finding techniques that allow us to automatically reason about other desired properties of e-voting systems such as election verifiability. Moreover, a formal specification of system mechanisms that were abstracted away in this paper such as spoil-audit and risk-limiting audit for verifiability purposes can be pursued as future work.

\paragraph{\bf{Acknowledgments.}} This work was carried out under the NSF-funded Voting Systems Architectures for Security and Usability project. The principal author is also partly funded by TUBITAK. We would like to thank Ben Smyth, Olivier Pereira, and Thomas Gibson-Robinson for their helpful technical discussions.


\bibliographystyle{plain} 
\bibliography{References}
\end{document}